\documentclass[aps,prl,twocolumn,amsmath,amssymb,nofootinbib,superscriptaddress,titlepage]{revtex4}
\usepackage{graphicx}
\usepackage{epstopdf}
\usepackage{graphicx}
\usepackage{dcolumn}
\usepackage{bm}
\usepackage{epsf}
\usepackage{amsfonts}

\newcommand{\be}{\begin{equation}}
\newcommand{\ee}{\end{equation}}
\newcommand{\beqn}{\begin{eqnarray}}
\newcommand{\eeqn}{\end{eqnarray}}

\begin{document}

\title{Multiscale photosynthetic exciton transfer}

\author{A.K. Ringsmuth}
\email{a.ringsmuth@uq.edu.au} 
\affiliation{Institute for Molecular Bioscience, University of Queensland, St. Lucia, QLD
4072, Australia}
\affiliation{Centre for Engineered Quantum Systems, University of Queensland, St. Lucia, QLD
4072, Australia}
\author{G.J. Milburn}
\affiliation{Centre for Engineered Quantum Systems, University of Queensland, St. Lucia, QLD
4072, Australia} 
\author{T.M. Stace}
\affiliation{Centre for Engineered Quantum Systems, University of Queensland, St. Lucia, QLD
4072, Australia}

\date{March 17, 2011}

\maketitle

\textbf{Photosynthetic light harvesting provides a natural blueprint for bioengineered and biomimetic solar energy and light detection technologies. Recent evidence \cite{engel07, collini10, panit10, schlau09, lee07, sarovar10, ishizaki09, ishizaki09(2), mohseni08, rebentrost09, chin10} suggests some individual light harvesting protein complexes (LHCs) and LHC subunits efficiently transfer excitons towards chemical reaction centers (RCs) via an interplay between excitonic quantum coherence, resonant protein vibrations, and thermal decoherence. The role of coherence \textit{in vivo} is unclear however, where excitons are transferred through multi-LHC/RC aggregates over distances typically large compared with intra-LHC scales \cite{fassioli09, barzda96, broess06}. Here we assess the possibility of long-range coherent transfer in a simple chromophore network with disordered site and transfer coupling energies. Through renormalization we find that, surprisingly, decoherence is diminished at larger scales, and long-range coherence is facilitated by chromophoric clustering. Conversely, static disorder in the site energies grows with length scale, forcing localization. Our results suggest sustained coherent exciton transfer may be possible over distances large compared with nearest-neighbour (n-n) chromophore separations, at physiological temperatures, in a clustered network with small static disorder. This may support findings suggesting long-range coherence in algal chloroplasts \cite{collini10}, and provides a framework for engineering large chromophore or quantum dot high-temperature exciton transfer networks.}

Individual LHCs exploit a balance of quantum coherence and thermal decoherence to improve their exciton transfer efficiency. At intra-LHC scales, vibrational resonances in the surrounding protein help to preserve excitonic coherence when electronic transfer couplings and protein reorganization energies converge \cite{lee07, ishizaki09(2), ishizaki09, rebentrost09(2)}. Thermal dephasing noise cooperates with this coherence to transfer excitons through the network by suppressing transfer-inhibiting interference effects and exploiting spectral broadening of site energies to overcome Anderson localization; the result is transfer efficiency exceeding that possible through purely coherent or purely incoherent mechanisms \cite{rebentrost09, chin10}. An interesting question is whether these effects can improve physiologically relevant, long-range transfer efficiency through multi-LHC/RC networks and, if so, whether this enhancement is due only to accumulated intra-LHC enhancements or some more subtle long-range mechanism. 

\begin{figure}
\includegraphics[angle=0,width=1\columnwidth]{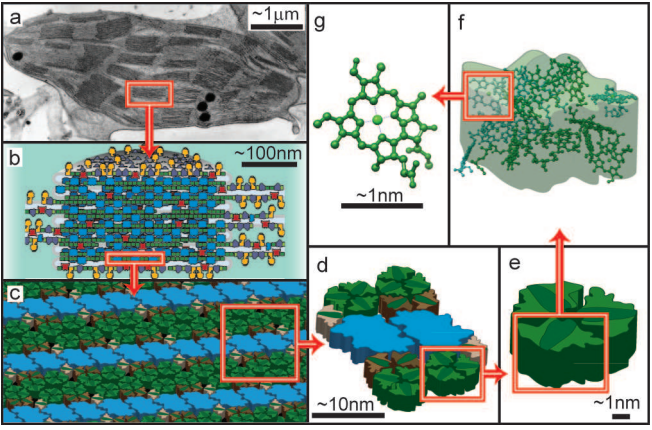}
\caption{\label{fig:multiscale} \textbf{Structural heirarchy of the light harvesting machinery from higher plants.} a) Transmission electron micrograph of a sugarcane chloroplast, fixed, sectioned and stained. Densely stacked grana regions of the thylakoid membrane are visible. Image courtesy of Prof. Robert Birch, used with permission. b) Pseudo-3D cross-section of a granal membrane stack showing embedded aggregates of photosystem II-light harvesting complex II (PSII-LHCII) supercomplexes interspersed with other complexes. c) 3D schematic of membrane-embedded semicrystalline aggregate of PSII-LHCII supercomplexes. d) PSII-LHCII supercomplex containing LHCs (green, brown) coupled to RC-containing core complexes (blue). e) Trimeric LHCII. f) Monomeric subunit of LHCII, with crystal structure overlaid. This complex binds 14 chlorophyll (8 chlorophyll-a, 6 chlorophyll-b) and 4 carotenoid chromophores. g) Chlorophyll-a chromophore.}
\end{figure}

An arbitrary chromophore network with disordered coupling topology may be considered as a hierarchy of nested clusters, where a cluster is defined as a group of sites coupled more strongly to each other than to sites outside the cluster, and the degree of clustering is quantified by the ratio of intracluster to intercluster coupling energies. At the smallest scale are clusters of strongly coupled chromophores, such as monomeric LHC subunits (Figure \ref{fig:multiscale} f). At the next scale, each chromophore cluster forms an effective site, which itself clusters with other similar sites; monomeric LHC subunits cluster to form whole LHCs. The hierarchy may proceed further, to LHC aggregates and/or LHC-RC supercomplexes, to aggregates of supercomplexes, which in some cases finally cluster in 3D membrane stacks called grana \cite{dekker05}. The efficiency of transfer to RCs depends strongly on structure at multiple scales within this hierarchy. Within some species, organization of LHC aggregates can facilitate functional switching between a high transfer efficiency state and a so-called nonphotochemical quenching state, which dissipates excess excitons as heat \cite{horton08}. However, mechanisms of this switch remain controversial and their relationship to fundamental exciton transfer mechanisms is yet to be fully understood. Chromophoric clustering hierarchies also exist in some artificial light harvesting materials such as organic semiconductors \cite{spano09, collini09}.   
%%%%%%%%%%%%%%%%%%%%%%%%%%%%%%%%%%%%%%%%%%%%%%%%%%%%%%%%%%%%%%%%%%%%%%%%%%%%%%%%%%%%%%%%%%%%%%%
\subsection{Dimer}\label{dimer}
Initially we study the crossover from coherent (tunnelling) to incoherent (hopping) energy transfer in a pair (dimer) of undriven chromophores with separation $d=|\textbf{d}|=|\textbf{r}^{(2)}-\textbf{r}^{(1)}|$. This dimer is our basic clustering unit, loosely analogous to a LHC subunit. We assume for the complete Hamiltonian \cite{mahan00},
\be
H ={H^S} + H^B + {H^{SB}}, \nonumber
\ee
where $H^S$ is a spin-boson type, two-site transfer Hamiltonian with site energies $E^{(1,2)}$, static disorder $\epsilon= E^{(1)}-E^{(2)}$, and transfer coupling $\Delta$. The bath Hamiltonian $H^B=\hbar \sum_\textbf{q} \omega(\textbf{q}) a(\textbf{q})^\dagger a(\textbf{q})$ is the sum over harmonic vibrational modes with wave vectors $\textbf{q}$ and frequencies $\omega(\textbf{q})$. The system-bath interaction is $H^{SB} = i\sum_\textbf{q} \hat{G}^{(12)}(\textbf{q}) \left(a(\textbf{q})-{a(-\textbf{q})}^\dagger\right)$, where $\hat{G}^{(12)}(\textbf{q})$ is an effective exciton-phonon coupling operator for the dimer (see Supplementary Information).

We derive the system-bath coupling, $g^{(12)}(\textbf{q})$, spectral density and a Markovian master equation for ${\rho_I}$, the combined density matrix of the site populations in an interaction picture with respect to $H^S + H^B$ (see Supplementary Information). The spectral density, \mbox{$J(\omega)=2\pi\sum_\textbf{q} \delta(\omega-\omega(\textbf{q}))|{g^{(12)}(\textbf{q})|^2} $}, which quantifies the strength of the system-bath coupling, is
\be
J(\omega)=  B\left(\omega^3-\frac{\omega^2v}{d}\sin\left(\frac{\omega d}{v}\right)\right),\label{spec0}
\ee
where $B=\hbar {D}^2/(4\pi\mu v^5)$ with $D$ the exciton-phonon deformation potential, $\mu$ the mass density and $v$ the speed of sound.  In the ordered case ($\epsilon=0$), the site populations of an initially localized exciton show coherent tunnelling between sites at frequency $2\Delta$, damped at temperature-dependent decoherence rate $F(2\Delta)$. We call the under-damped regime, where quantum tunelling dominates, `coherent', and the critically- and over-damped regime, where thermal dephasing and hopping dominate, `incoherent'. The two regimes are defined as
\be
\mbox{Coherent:}~F(2\Delta)< 2\Delta, ~~~\mbox{Incoherent:}~F(2\Delta)\geq 2\Delta. \label{cri}
\ee 
%%%%%%%%%%%%%%%%%%%%%%%%%%%%%%%%%%%%%%%%%%%%%%%%%%%%%%%%%%%%%%%%%%%%%%%%%%%%%%%%%%%%%%
\subsection{Clustered network and renormalization} \label{tetramer}
\begin{figure*}
\includegraphics[angle=0,width=1\textwidth]{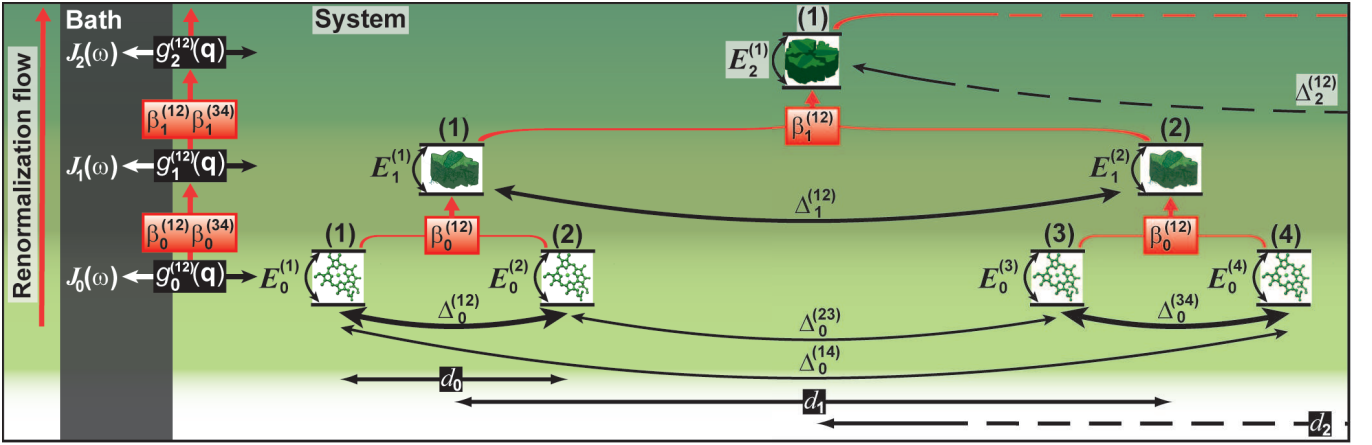}
\caption{\label{fig:tetramer} \textbf{Renormalization flow of dimeric chromophore network hierarchy.} A tetramer at the 0th level is coarse-grained to an effective dimer at the 1st level via the operator $\beta_k^{(ij)}$. This dimer is paired with a similar effective dimer and the coarse-graining repeated to obtain an effective dimer at the 2nd level, and so on. The site energies $E_k^{(i)}$, transfer coupling energies $\Delta_k^{(ij)}$ and system-bath coupling energies $g_k^{(12)}$ are transformed between levels. At each level, analogous structures from the higher plant photosynthetic structural hierarchy are presented in parallel with the dimeric network hierarchy: 0) Chlorophyll-a, 1) LHCII monomer, 2) LHCII trimer.}
\end{figure*}
Consider a network of coupled dimers extending to length scales large compared with the intradimer separation, analogous to an aggregate of elementary chromophore clusters such as LHC subunits (Figure \ref{fig:tetramer}). Our treatment is extensible to an arbitrary network geometry but we assume a one-dimensional arrangement, symmetric about the origin, with uniform intradimer separations $d_0$ and interdimer (centre-centre) separations $d_1$ such that $d_1>2d_0$. This forms the 0th level of the structural hierarchy, which we now analyze through renormalization. We relax the assumption prevalent in earlier studies \cite{jang04, fetisova04, fassioli09, broess06}, that there is some length scale at which tunnelling becomes unimportant. Instead, we consider the dynamics at higher hierarchy levels as perturbations to the dynamics at lower levels \cite{wilson75}, allowing for the possibility of coherent transfer at all scales. To assess the dynamics at each level, we formulate an iterable renormalization procedure by which a cluster of $n$ sites at the $k$th level is mapped to a single effective site at the $(k+1)$th level, located at $\textbf{r}_{k+1}=\frac{1}{n} \sum_{j=1}^n \textbf{r}^{(j)}_k$ (See Supplementary Information). This is achieved via a coarse-graining operator which maps the $n$ single-excitation electronic eigenstates (excitons) of the cluster to the unique single-excitation eigenstate of the effective site. 

Figure \ref{fig:tetramer} shows the renormalization flow for hierarchy levels 0-2. The flow is characterized by transformations of the site, transfer coupling and system-bath coupling energies between levels. At the $k$th level, the excitation energy of a site is approximately the average of its 0th-level intracluster site energies and the transfer coupling between sites is approximately the average of the 0th-level intercluster transfer couplings (see Supplementary Equations (11) and (13). The system-bath coupling at the $k$th level is given by the exciton-phonon coupling strength, $M(\textbf{q})$, modulated by a product of periodic terms dependent on the network geometry according to
\be
g_{k}^{(12)}(\textbf{q})= M(\textbf{q})\sin{\left(\frac{\textbf{d}_{k}\cdot\textbf{q}}{2}\right)}\prod_{j=0}^{k-1}\cos{\left(\frac{\textbf{d}_j\cdot\textbf{q}}{2}\right)}.~~~~\label{grenormgen}
\ee
The spectral density for an effective dimer at the $k$th ($k\geq 1$) hierarchy level, $J_k(\omega)$, is found by letting ${g^{(12)}(\textbf{q}})\rightarrow g_k^{(12)}(\textbf{q})$ in the spectral density and substituting equation (\ref{grenormgen}). This yields 
\beqn
J_k(\omega) &=&B W_k(\omega)\omega^3,\label{specgen}
\eeqn
where $W_k(\omega)$ is a modulating function that depends on the network geometry and characterizes the changes in the bath's coupling to the network at different levels (see Supplementary Equation (10)). The coefficients $A_{i}(\omega)=\omega d_i/v$, with $d_i = 2^{(i-1)}d_1~(i\geq1)$ the intradimer separation at the $i$th level. The effect of the modulating function $W_k(\omega)$ is to filter the system-bath interaction such that the phonon wavelengths which strongly affect the network at a given level are proportional to the length scales characteristic of that level. At sufficiently high levels, the transfer coupling is diminished such that the small-frequency limit $\omega\ll v/d_k$ holds and $J_k(\omega)\approx B d_k^2\omega^5/(6v^2)$, from which it follows that 
\be
J_{k+1}(\omega)\approx \left(\frac{d_{k+1}}{d_k}\right)^2J_{k}(\omega)\label{specap}. 
\ee
In this limit, given our chosen network geometry, $J_{k+1}(\omega)\approx 4 J_{k}(\omega)$ for $k\geq1$.
%%%%%%%%%%%%%%%%%%%%%%%%%%%%%%%%%%%%%%%%%%%%%%%%%%%%%%%%%%%%%%%%%%%%%%
\subsection{Scale-based crossover}
\begin{figure}
\includegraphics[angle=0,width=0.85\columnwidth]{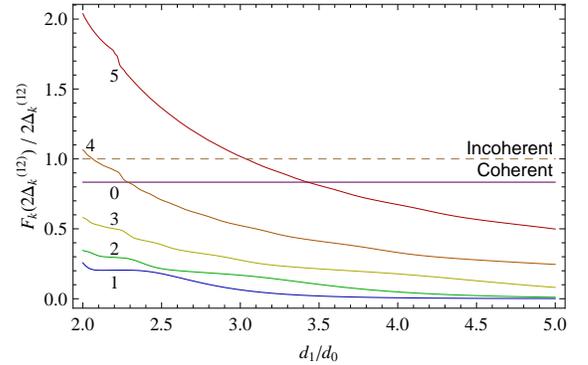}
\caption{\label{fig:plots} \textbf{Scale-based crossover simulation results.} Shown for first six hierarchy levels (curves labelled) at T=293K in ordered ($\epsilon_0^{(ij)}=0$) networks with various degrees of chromophoric clustering ($d_1/d_0$). Dashed line separates coherent (below) and incoherent (above) dynamical regimes.}
\end{figure}

We assess the transfer dynamics at the $k$th level in an ordered network ($\epsilon_0^{(ij)}=0$) %, using equations (\ref{specgen}) and (\ref{deltagen}) to compute
by calculating the ratio of the decoherence rate to the transfer coupling frequency, $F_k(2\Delta_k)/(2\Delta_k)$ where $F_k(2\Delta_k)= J_k(2\Delta_k)(N(2\Delta_k)+\frac{1}{2})$, where $~N(\omega)=(e^{\hbar \omega/k_BT}-1)^{-1}$ is the phonon occupation of the bath at temperature $T$, and compare this with definitions in equation (\ref{cri}). Figure \ref{fig:plots} plots this ratio for the first six levels in ordered networks with varying degrees of clustering ($d_1/d_0>2$). The parameter set used to generate figure \ref{fig:plots}, chosen for biological relevance (see Supplementary Information), supports coherent transfer at hierarchy levels 0-3 for all degrees of clustering, since $F_{k}(2\Delta_{k})/(2\Delta_{k})<1$. However, at levels 4 and 5 transfer is coherent only in more clustered networks, suggesting that clustering facilitates coherent transfer rather than inhibiting it, as may be assumed. Although decoherence dominates at intermediate hierarchy levels with low degrees of clustering, at high levels (where $2\Delta_k\ll v/d_k$), $F_{k+1}(2\Delta_{k+1})/(2\Delta_{k+1})\approx4\left(\Delta_{k+1}/\Delta_k\right)^3 F_{k}(2\Delta_{k})/(2\Delta_{k})$, and therefore at sufficiently high levels, $F_{k}(2\Delta_{k})/(2\Delta_{k})$ vanishes. This indicates that in a large, ordered network, coherent transfer is supported at the largest length scales accessible within the excitation lifetime, provided it is also supported at all smaller scales. The latter requirement is met when the transfer coupling occupies the small frequency regime at all length scales. 

We interpret these results in the following way, assuming rapid transfer with sustained coherence may be beneficial to photosynthetic efficiency \cite{engel07, collini10, panit10, schlau09, lee07, sarovar10, ishizaki09, ishizaki09(2), mohseni08, rebentrost09, chin10}. In a biologically relevant parameter space of our network, rapid coherent oscillations dominate at small scales but are quickly damped. While coherence is more sustained at larger scales, the transfer rate is disadvantageously slow. Chromophoric clustering balances these two extremes at intermediate scales by restricting the bath modes able to couple to the network, thereby reducing the decoherence rate relative to the transfer rate. Consequently, coherent transfer is supported at larger length scales than in an equivalent unclustered network. 
%%%%%%%%%%%%%%%%%%%%%%%%%%%%%%%%%%%%%%%%%%%%%%%%%%%%%%%%%%%%%%%%%%%%%%%%%%%%%%%%%%%%
\subsection{Static disorder forces localization}
We now introduce intradimeric static disorder at the $0$th hierarchy level, $\epsilon_0^{(ij)}=E_0^{(i)}-E_0^{(j)}$ ($i$ odd, $j=i+1$), sampled from a Gaussian distribution with variance $\sigma^2(\epsilon_0^{(ij)})$. Since Gaussian variances add directly, the variance and mean of the static disorder at the $k$th hierarchy level respectively are 
\be
\sigma^2(\epsilon_k^{(ij)})=2^k\sigma^2(\epsilon_0^{(ij)}),\quad
\bar{\epsilon}_k^{~(ij)}=\sqrt{\frac{2^{k-1}}{\pi}}~\sigma(\epsilon_0^{(ij)}).\nonumber%\label{mean_k}.
\ee
The mean intradimer static disorder increases with $k$. At some level,  $k_{A}$, this disorder will exceed the tunnelling rate (which decreases with $k$), 
i.e.\ $\bar{\epsilon}_{k_A}^{~(ij)}\gg\Delta_{k_A}^{(ij)}$, so the excitons become localized on a single effective site. This is a manifestation of Anderson localization. Previous studies \cite{rebentrost09, chin10} have found that at the 0th hierarchy level, spectral broadening due to thermal dynamic disorder can overcome Anderson localization to facilitate efficient transfer. However, we have shown that at higher hierarchy levels, dynamic disorder is reduced while static disorder is increased. Anderson localization therefore provides the ultimate limit to coherent  long-range exciton transfer.  Transfer over scales longer than ${d}_{k_A}$ must be by thermally induced hopping.
%%%%%%%%%%%%%%%%%%%%%%%%%%%%%%%%%%%%%%%%%%%%%%%%%%%%%%%%%%%%%%%%%%%%%%%%%%%%%%%%%%%%%%%%%

\subsection{Conclusions}
We have analyzed scaling trends in exciton transfer in a chromophore network with self-similar, hierarchical, dimeric clustering geometry, and with Markovian system-bath interaction. We have assessed bath-induced crossover from coherent to incoherent transfer across the hierarchy's spectrum of length scales via a novel renomalization procedure. For networks without static disorder in the site energies, the transfer coupling varies inversely with both hierarchy level and degree of chromophore clustering. Counterintuitively, excitonic decoherence also varies inversely with both level and clustering. In a biologically relevant parameter space, our chosen network geometry supports coherent transfer for levels 0-3, and also higher levels in more clustered networks. Chromophore clustering therefore facilitates long-range coherence rather than inhibiting it. This may suggest that the clustering hierarchies ubiquitous in biological chromophore networks help exploit excitonic coherence over scales larger than achievable in equivalent unclustered networks. 

Anderson localization due to static disorder provides the ultimate limit on coherent long-range transfer, rather than thermal dynamic disorder or large separations between chromophore clusters. This suggests that determining the significance of coherence-assisted transfer over physiologically-relevant length scales in real systems may require a focus on the scaling of static disorder, for example by studying inhomogeneous broadening. Beyond biological systems, our results suggest the possibility of long-range coherence-assisted transfer at ambient temperatures in a chromophore network engineered with small static disorder and clustering geometry which minimizes decoherence at larger length scales. This may also hint at a more general principle concerning networks of electronically coupled quantum systems in condensed matter: Hierarchically clustered network geometries, carefully designed to complement the dielectric and mechanical properties of their surrounding media, may aid in sustaining long-range coherent interactions at high temperatures. This is potentially of interest for quantum information applications. 

Avenues for extension of our model include generalizing the dynamical treatment to strong system-bath coupling at low hierarchy levels; reference \cite{rebentrost09(2)} and references therein describe such non-Markovian treatments. We speculate that including dissipative processes and analyzing their scaling may be useful in studying nonphotochemical exciton quenching mechanisms, in which long-range LHC clustering geometry is implicated \cite{horton08}. Finally, it may be of interest to generalize the renormalization procedure to geometries which are higher-dimensional, have non-uniform site separations and/or have variations in cluster size between hierarchy levels, and to accommodate nonlinear phonon dispersion due to material differences between hierarchy levels. 

%%%%%%%%%%%%%%%%%%%%%%%%%%%%%%%%%%%%%%%%%%%%%%%%%%%%%%%%%%%%5
\section{Supplementary Information}

\subsection{Model}
We assume for the complete Hamiltonian \cite{mahan00},
\beqn
H &=&{H^S} + H^B + {H^{SB}},~~\text{where} \setcounter{equation}{1}\label{Hd}\\
H^S &=& E^{(1)} |1\rangle \langle 1|+E^{(2)} |2\rangle \langle 2|-\Delta(|1\rangle \langle 2|+|2\rangle \langle 1|),~~~\label{HSd}\\ 
H^B &=& \hbar \sum_\textbf{q} \omega(\textbf{q}) a(\textbf{q})^\dagger a(\textbf{q}),\label{HBd}\\
H^{SB} &=& i\sum_\textbf{q} \hat{G}^{(12)}(\textbf{q}) \left(a(\textbf{q})-{a(-\textbf{q})}^\dagger\right).\label{HSBd}
\eeqn
Here, $\{|1\rangle,|2\rangle\}$ is the single-excitation basis for two sites with energies $E^{(1,2)}$, static disorder $\epsilon= E^{(1)}-E^{(2)}$, and transfer coupling $\Delta$. $a(\textbf{q})$ annihilates a phonon with wave vector $\textbf{q}$, $\hat{G}^{(12)}(\textbf{q})= M(\textbf{q}){\hat{\rho}}(\textbf{q})$ is an effective exciton-phonon coupling operator for the dimer.
\subsection{System-bath coupling and spectral density}
The spectral density depends on the microscopic system-bath coupling, ${g^{(12)}(\textbf{q})}$. Details of system-bath couplings for biological LHCs are not yet well known \cite{chin10}, but are nonetheless generically described \cite{mahan00} by an effective exciton-phonon coupling operator, $\hat{G}^{(12)}(\textbf{q})= M(\textbf{q}){\hat{\rho}}(\textbf{q})$, with $M(\textbf{q})$ the coupling strength and    
\beqn
{\hat{\rho}}(\textbf{q})&=&\sum_{m,n}\left(\int{d^3\textbf{r}~{\psi^{(m)}_{X}}^*(\textbf{r})\psi^{(n)}_{X}(\textbf{r}) e^{-i\textbf{q}\cdot \textbf{r}}}\right)c_m^\dagger c_n~~\nonumber\\
&=&\mathcal{P}(\textbf{q})\sum_{m} e^{-i\textbf{q}\cdot \textbf{r}_{m}}~|m\rangle\langle m| \label{density0}
\eeqn
the Fourier transform of the exciton density operator. $\psi^{(i)}_{X}(\textbf{r})$ is the wave function for an excitation on site $i$, and $\mathcal{P}(\textbf{q})= \int d^3\textbf{r}|\psi_{X}(\textbf{r})|^2e^{-i\textbf{q}\cdot\textbf{r}}$ is the form factor of the wave function \cite{stace05}. We assume single-site excitations are localized within length $l$ much smaller than the intradimer separation $d$, such that $\mathcal{P}(\textbf{q})\approx 1$ for all $|\textbf{q}|$ values of interest. Evaluating equation (\ref{density0}) for the dimer comprising sites 1 and 2 gives, 
\beqn
{\hat{\rho}}(\textbf{q})&=&\cos\left(\frac{\textbf{d}\cdot\textbf{q}}{2}\right)\left(|1\rangle\langle 1|+|2\rangle\langle 2|\right)\nonumber\\
&~&~~+i\sin\left(\frac{\textbf{d}\cdot\textbf{q}}{2}\right)\left(|1\rangle\langle 1|-|2\rangle\langle 2|\right)\nonumber\\
&\equiv& \sin\left(\frac{\textbf{d}\cdot\textbf{q}}{2}\right)\hat{Z},\label{dimercoup}\nonumber
\eeqn
where $\hat{Z}=|1\rangle\langle 1|-|2\rangle\langle 2|$. The term proportional to the identity is ignored since it only shifts the phonon energies, and the phase factor is also ignored. Then we can write $\hat{G}^{(12)}(\textbf{q})=g^{(12)}(\textbf{q})~\hat{Z}$, where
\be
g^{(12)}(\textbf{q})=M(\textbf{q})\sin\left(\frac{\textbf{d}\cdot\textbf{q}}{2}\right).\label{sbcoup0}
\ee
The system-bath interaction Hamiltonian is therefore
\be
{H^{SB}} =\hat{Z}\sum_\textbf{q}g^{(12)}(\textbf{q})\left(a(\textbf{q})-{a(-\textbf{q})}^\dagger\right).\label{HSB121_}
\ee
In order to compute the spectral density, we assume as a first approximation, deformation coupling of the form $M(\textbf{q})=D|\textbf{q}|\sqrt{\hbar/{2\mu V \omega(\textbf{q})}}$, which accurately describes a quantum dot in a crystal \cite{mahan00}. Here, $\mu$ is the mass density, $D$ the deformation potential, and $V$ the quantization volume. We stress that the precise form of this coupling is used as a guide only, and the salient features of our model on which we base our conclusions are independent of the specific parameters therein.  The spectral density at the $0$th level is computed by assuming linear phonon dispersion in the bath and approximating with the integral,
\beqn
J(\omega) =\frac{V}{4\pi^2}\int dq \,d\phi\, d\theta\, q^2\sin{\phi}~\delta(\omega-v q)\label{spec0int}
|g^{(12)}_{q,\phi,\theta}|^2  \label{specint},
\eeqn
which, using equation (\ref{sbcoup0}), evaluates to equation (1) in the main text.

\subsection{Dimer master equation and dynamics}
Here we derive a non-dissipative Markovian master equation for $\rho_I$, the combined density matrix of the dimer site populations in an interaction picture with respect to ${H^S}+H^B$. Non-Markovian effects enhance the preservation of excitonic coherence at intra-LHC scales, compared with the predictions of Markovian models \cite{ishizaki09(3), ishizaki09(2), rebentrost09(2), sarovar10, chin10}. However, the ultimate focus of this analysis is long-range transfer, which we show to be less affected by system-bath interactions than short-range transfer, and therefore more appropriate for a Markovian description. Moreover, a Markovian treatment at intra-LHC scales is a conservative choice since it predicts faster decoherence than non-Markovian treatments \cite{ishizaki09(2), rebentrost09(2)}. 

Transforming equation (\ref{HSB121_}) into an interaction picture with respect to ${H^S}+H^B$, in which $a(\textbf{q},t)=a(\textbf{q})e^{-i\omega(\textbf{q}) t}$, we find
\begin{equation}
\hat{Z}(t) = \sum_{\omega '\in \{0,\phi\}} P(\omega ') e^{-i \omega ' t} + {P(\omega ')^\dagger e^{i\omega ' t},}\nonumber
\end{equation}
where $\phi=\sqrt{4\Delta^2+\epsilon^2}$. The operators $P(0)=\frac{\epsilon\Delta}{\phi^2}\left(-{\hat{\sigma}_+}-{\hat{\sigma}_-}+\frac{\epsilon(\epsilon+\phi)}{2\Delta\phi^3}{\hat{Z}}-\frac{4\Delta}{\phi}\mathbb{I}\right)$ and $P(\phi)=\frac{\Delta}{\phi^2}\left((\epsilon-\phi){\hat{\sigma}_+}+(\epsilon+\phi){\hat{\sigma}_-}+2\Delta{\hat{Z}}\right)$, where $\hat{\sigma}_+=|1\rangle\langle 2|,~\hat{\sigma}_-=|2\rangle\langle 1|,~\hat{Z}=|1\rangle\langle 1|-|2\rangle\langle 2|$. The terms proportional to $\hat{Z}$ cause pure dephasing in the site basis while those proportional to $\hat{\sigma}_+$ or $\hat{\sigma}_-$ induce transitions between sites. The term proportional to the identity can be ignored, giving the interaction picture Hamiltonian,
\beqn 
{H_I}^{SB}(t)=\hat{Z}(t) \sum_\textbf{q} g^{(12)}(\textbf{q})\left(a(\textbf{q})e^{-i\omega(\textbf{q}) t} +a(\textbf{q})^\dagger e^{i\omega(\textbf{q})t}\right).\nonumber
\eeqn
Integrating over the von Neumann equation for the density matrix, $W$, of the total system then tracing out the phonon modes gives for the combined density matrix of the site populations,
\be
\dot\rho_I(t)= -\int_{t_0}^t dt'{Tr_{ph}}\{[H_I^{SB}(t),[H_I^{SB}(t'),{W_I}(t')]]\}.\nonumber
\ee
We assume weak system-bath coupling and rapid bath relaxation, allowing a Born-Markov approximation to be made by setting $t_0=-\infty$ and replacing $W_I(t')\rightarrow W_I(t)$. The result is
\beqn
\dot\rho_I = \sum_{\omega '=0,\phi} J(\omega ')((N(\omega')+1)\mathcal{D}[P(\omega ')]\rho_I \nonumber\\
+ N(\omega ')\mathcal{D}[{P(\omega ')}^\dagger]\rho_I ),\label{master}\nonumber
\eeqn
where $\mathcal{D}[A]\rho \equiv A\rho A^\dagger - \frac{1}{2}(A^\dagger A\rho + \rho A^\dagger A)$. $~N(\omega)=(e^{\hbar \omega/k_BT}-1)^{-1}$ is the phonon occupation of the bath at temperature $T$ and $J(\omega)$ is the spectral density. Exciton recombination processes are not included, since we assume the lifetime to be long compared with the time required for transfer to a RC, which is necessary for the near-unity quantum efficiencies of unstressed photosynthetic membranes. 

In the ordered case ($\epsilon=0$), an initially localized excitation (${\rho_I}_{1,1}(0)=1$) evolves in the Schr\"odinger picture as
\be
{\rho_S}_{j,j}(t)=\frac{1}{2}\left(1\pm e^{-F(2\Delta)t}\cos(2\Delta t)\right),~~j=1,2, \label{schroedinger_soln}\nonumber
\ee
showing coherent tunnelling between sites at frequency $2\Delta$%($\phi$ when $\epsilon \neq 0$)
, damped by bath-induced decoherence at rate $F(2\Delta)= J(2\Delta)(N(2\Delta)+\frac{1}{2})$, where $~N(\omega)=(e^{\hbar \omega/k_BT}-1)^{-1}$ is the phonon occupation of the bath at temperature $T$. 

\subsection{Renormalization procedure}
Renormalization from the $k$th to the $(k+1)$th hierarchy level is achieved via a coarse-graining operator which maps the $n$ single-excitation electronic eigenstates (excitons) of a chromophore cluster at the $k$th level to the unique single-excitation eigenstate of the equivalent effective site at the $(k+1)$th level. To construct the operator, we consider a pair of dimers (tetramer)(Figure 2 in the main text) for which the Hamiltonians (\ref{Hd}), (\ref{HSd}) and (\ref{HSBd}) are generalized as
\beqn 
H_0^{(1..4)} =& &{H_0^S}^{(1..4)} + H^B + {H_0^{SB}}^{(1..4)},\nonumber\\
{H_0^S}^{(1..4)} =& &\sum_i E_0^{(i)} |i\rangle_0\langle i|-\frac{1}{2}\sum_{i,j}\Delta^{(ij)}_0(|i\rangle_0\langle j|+|j\rangle_0\langle i|)\nonumber\\
& & ~~~ \text{for} ~~i\neq j;~ i,j=1..4,\label{HS}\nonumber\\ 
{H_0^{SB}}^{(1..4)} =& &i\sum_q \hat{G}_0^{(1..4)}(\textbf{q}) \left(a(\textbf{q})-{a(-\textbf{q})}^\dagger\right),\label{HSB}\nonumber
\eeqn
with numerical subscripts labeling hierarchy level. The bath Hamiltonian $H^B$ is still given by equation (\ref{HBd}). 
For a dimer comprising sites $i$ and $j$ at the $k$th hierarchy level, with effective static disorder $\epsilon_k^{(ij)}=E_k^{(i)}-E_k^{(j)}$ and transfer coupling $\Delta_k^{(ij)}$, the eigenstates are $|\pm\rangle_k^{(ij)}=\cos{{\theta_k}_\pm^{(ij)}}|i\rangle_k+\sin{{\theta_k}_\pm^{(ij)}}|j\rangle_k)$, where ${\theta_k}_\pm^{(ij)}=\arccos{\left(-(\epsilon_k^{(ij)}\pm\phi_k^{(ij)})/\sqrt{4{\Delta_k^{(ij)}}^2+(\epsilon_k^{(ij)}\pm\phi_k^{(ij)})^2}\right)}$, with eigenenergies $\mathcal{E}^{(ij)}_{k_\pm}=\frac{1}{2}(E_k^{(i)}+E_k^{(j)}\pm\phi_k^{(ij)})$. Here, $\phi_k^{(ij)}=\sqrt{4{\Delta_k^{(ij)}}^2+{\epsilon_k^{(ij)}}^2}$. The interdimer transfer dynamics are dominated by the low-energy eigenstates of the tetramer. Therefore, we define a renormalization operator for the dimer, which projects onto its lowest energy eigenstate. For $j,~n$ even and $i=j-1,~m=n-1$, 
\be
\beta_k^{(ij)} =  |j/2\rangle_{k+1}\langle -|^{(ij)}_k+ \sum_{n\neq j}|vac\rangle^{(n/2)}_{k+1}\left(\langle vac|^{(m)}_k+\langle vac|^{(n)}_k\right).\label{beta}
\ee
The complete Hamiltonian for a dimer at the $(k+1)$th level is obtained by renormalizing the corresponding tetramer at the $k$th level:
\beqn
H_{k+1}^{(12)} &=& \beta_k^{(12)}\beta_k^{(34)}{H_k^{(1..4)}}~{\beta_k^{(12)}}^\dagger{\beta_k^{(34)}}^\dagger \nonumber\\ 
&=&{H_{k+1}^S}^{(12)} + H^B + {H_{k+1}^{SB}}^{(12)},~~\text{where} \label{Hk+1}\nonumber\\
{H_{k+1}^S}^{(12)} &=& E_{k+1}^{(1)} |1\rangle_{k+1}\langle 1|+E_{k+1}^{(2)} |2\rangle_{k+1}\langle 2|\nonumber\\& &~~~~~~-\Delta^{(12)}_{k+1}(|1\rangle_{k+1}\langle 2|+|2\rangle_{k+1}\langle 1|),~~~ \label{Hk+1S}\nonumber\\
{H_{k+1}^{SB}}^{(12)} &=&\hat{Z}_{k+1}^{(12)}\sum_\textbf{q}g_{k+1}^{(12)}(\textbf{q})\left(a(\textbf{q})-a(-\textbf{q})^\dagger\right)\label{Hk+1SB}.\nonumber
\eeqn
$H_B$ commutes with $\beta_k^{(ij)}$; all levels of the hierarchy are coupled to the same bath, although the coupling, described by equation (3) in the main text, changes with $k$. Since the complete Hamiltonians at the $k$th and $(k+1)$th levels have the same form, the dynamics at both levels are described in the same way, with only the effective site, transition coupling and system-bath coupling energies being transformed between levels.  

\subsection{Renormalization of system-bath coupling and spectral density}
The spectral density is not renormalized directly but rather, at the $k$th hierarchy level it is computed from a renormalized system-bath coupling $g_{k}^{(12)}(\textbf{q})$. Equation (\ref{density0}) is renormalized using the operator defined in equation (\ref{beta}), letting the subscript $0\rightarrow k$ for generality:
\begin{align}
&{{\hat{\rho}}_{k+1}}({{\textbf{q}}} )=\beta_k^{(12)}\beta_k^{(34)}{{\hat{\rho}}}_{k}(\textbf{q} )~{\beta_k^{(12)}}^\dagger{\beta_k^{(34)}}^\dagger\nonumber\\
&=\cos\left(\frac{\textbf{d}_{k}\cdot\textbf{q}}{2}\right)\times\nonumber\\
~~~&\left(\left(\exp\left(i\textbf{d}_{k+1}\cdot\textbf{q}\right)\cos^2 {\theta_{k-}^{(12)}} +\cos^2 {\theta_{k-}^{(34)}}\right)\mathbb{I}_{k+1}\right.\nonumber\\
~~~&\left.+\left(\exp\left(i\textbf{d}_{k+1}\cdot\textbf{q}\right)\cos^2 {\theta_{k-}^{(12)}}-\cos^2 {\theta_{k-}^{(34)}}\right)\hat{Z}_{k+1}\right)\label{rhorenormgen}\nonumber\\
&\equiv\cos{\left(\frac{\textbf{d}_{k}\cdot\textbf{q}}{2}\right)}%\times\nonumber\\
%~~~&
\left[\cos{\left(\frac{\textbf{d}_{k+1}\cdot\textbf{q}}{2}\right)}\left(\cos^2 {\theta_{k-}^{(12)}}-\cos^2 {\theta_{k-}^{(34)}}\right)\right.\nonumber\\
~~~&\left.+\sin{\left(\frac{\textbf{d}_{k+1}\cdot\textbf{q}}{2}\right)}\left(\cos^2 {\theta_{k-}^{(12)}} +\cos^2 {\theta_{k-}^{(34)}}\right)\right]~\hat{Z}_{k+1}.%\label{rhorenorm}
\nonumber
\end{align}
Therefore, the system-bath coupling for the renormalized dimer at the $(k+1)$th level is given by equation (\ref{grenormgendis}). For an ordered network ($\epsilon_0^{(ij)}=0$), equation (\ref{grenormgendis}) reduces to
\be
g_{k+1}^{(12)}(\textbf{q})=M(\textbf{q})\cos{\left(\frac{\textbf{d}_{k}\cdot\textbf{q}}{2}\right)}\sin{\left(\frac{\textbf{d}_{k+1}\cdot\textbf{q}}{2}\right)}\label{grenorm},\nonumber
\ee
and recursive renormalization gives equation (3) in the main text. The spectral density at the $k$th ($k\geq 1$) hierarchy level, $J_k(\omega)$, is then found by letting ${g^{(12)}(q,\phi,\theta)}\rightarrow {g_k^{(12)}(q,\phi,\theta)}$ in equation (\ref{specint}) and substituting equation (3) from the main text. The result is equation (4) in the main text. In the small frequency limit where $\omega\ll v/d_k$, with $v$ the speed of sound, equation (5) in the main text allows for direct transformation between spectral densities at different hierarchy levels. Outside this limit, however, renormalization can proceed only via the system-bath coupling. This suggests that investigating multiscale exciton transfer in a specific LHC network or other real chromophore network may require the determination not only of the spectral density at some scale, but also of the exciton-phonon coupling parameters at the scale of individual chromophores. 

\setcounter{figure}{0}
\begin{figure}
\includegraphics[angle=0,width=0.9\columnwidth]{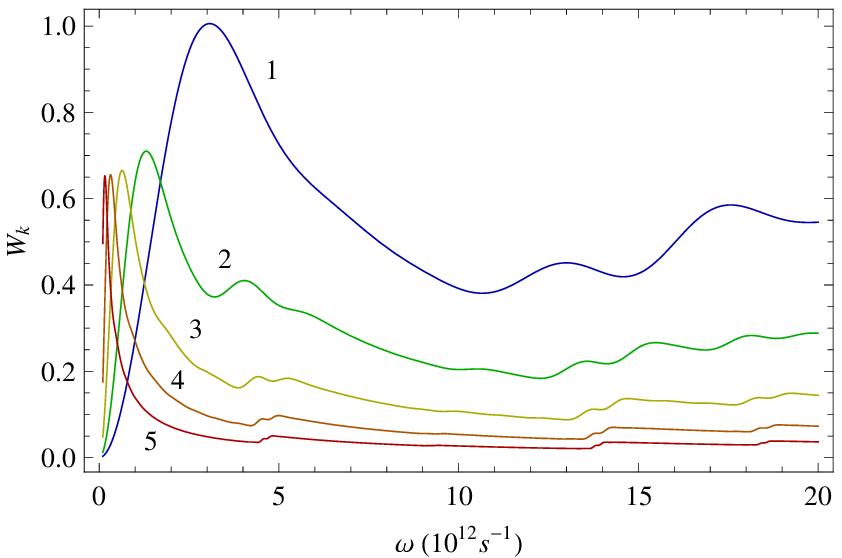}
\caption{\label{fig:modfun} \textbf{Spectral density modulating functions}, shown for first five renormalized hierarchy levels $k$ (curves labelled). Clustering is set at $d_1/d_0=3.4$, in analogy to trimeric LHCII.}
\end{figure}

\begin{figure}
\includegraphics[angle=0,width=0.9\columnwidth]{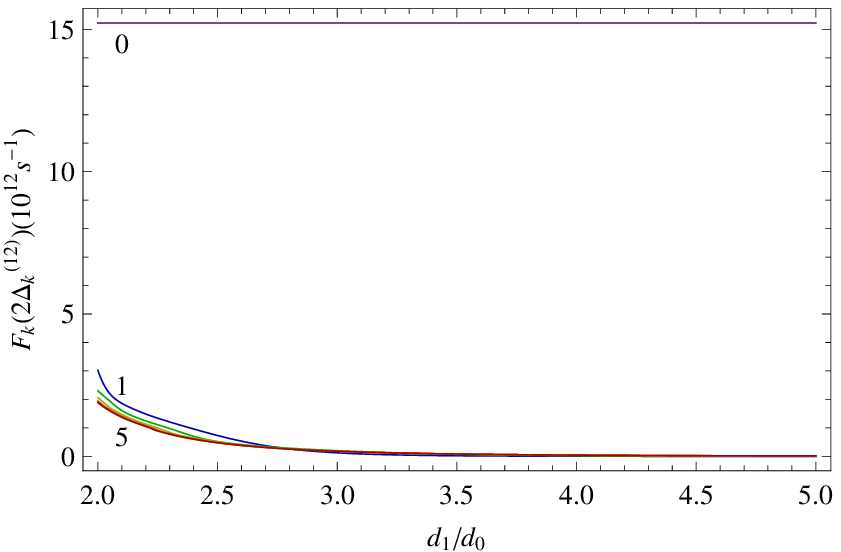}
\caption{\label{fig:deco} \textbf{Variations in the decoherence rate}, $F_{k}(2\Delta_k)$, with degree of network clustering ($d_1/d_0$), for first six hierarchy levels $k$ (curves labelled). Temperature, T=293K.}
\end{figure}

At higher levels of the hierarchy, the spectral density (equation (4) in the main text) contains a modulating function which depends on the network geometry:
\begin{eqnarray}
W_k(\omega)&=&\frac{1}{2^{k+1}}\int_{-1}^{1}dx [(1-\cos(A_k(\omega)x))\nonumber\\
%\nonumber\\&~&~~~~~~~~~~~~\times
&&\times\prod_{j=0}^{k-1}(1+\cos(A_{j}(\omega)x))].\label{modfun}
\end{eqnarray}
Figure \ref{fig:modfun} plots $W_k(\omega)$, for the first five renormalized levels, assuming $d_1/d_0=3.4$, in analogy to trimeric LHCII, the major LHC in higher plants, where the n-n intramonomer chromophore separation $d_0\approx10$\AA\ and the centre-centre intermonomer separation $d_1\approx34$\AA.

\subsection{Hierarchy energy transformations}
Through recursive renormalization we obtain exact expressions for the effective site, transition coupling and system-bath coupling energies at the $k$th ($k\geq1$) hierarchy level in terms of the $0$th level energies in an ordered network ($\epsilon_0^{(ij)}=0$):
\beqn
E_k^{(1)}&=&\frac{1}{2^k}\sum_{i=1}^{2^k}E_0^{(i)}-\sum_{i=0}^{k-1}\frac{1}{2^{k-i-1}}\sum_{m=1}^{2^{k-i-1}}\Delta_i^{(2m-1,2m)}\label{E1ex}\\
E_k^{(2)}&=&\frac{1}{2^k}\sum_{i=2^k+1}^{2^{k+1}}E_0^{(i)}\nonumber\\
&~&~~~~-\sum_{i=0}^{k-1}\frac{1}{2^{k-i-1}}\sum_{m=2^{k-i-1}+1}^{2^{k-i}}\Delta_i^{(2m-1,2m)}~~~~\label{E2ex}\\
\Delta_k^{(pq)}&=&\frac{1}{2^k}\sum_{i=2^k(p-1)+1}^{2^kp}\sum_{j=2^k(q-1)+1}^{2^kq}\Delta_0^{(ij)}\label{deltaex}\\
g_{k}^{(12)}&(\textbf{q})&= M(\textbf{q})\sin{\left(\frac{\textbf{d}_{k}\cdot\textbf{q}}{2}\right)}\prod_{j=0}^{k-1}\cos{\left(\frac{\textbf{d}_j\cdot\textbf{q}}{2}\right)}.\label{grenormex}
\eeqn
 Generalizing these transformations for a network with static disorder is nontrivial. Here we give transformations between the $k$th and $(k+1)$th hierarchy levels in a disordered network, which may be iterated to compute energies at an arbitrary level:
\beqn
E_{k+1}^{(1)}&=&E_k^{(1)}-\frac{4{\Delta_k^{(12)}}^2\phi_k^{(12)}}{4{\Delta_k^{(12)}}^2+(\epsilon_k^{(12)}-\phi_k^{(12)})^2},\label{renormsite1energy}\nonumber\\ E_{k+1}^{(2)}&=&E_k^{(3)}-\frac{4{\Delta_k^{(34)}}^2\phi_k^{(34)}}{4{\Delta_k^{(34)}}^2+(\epsilon_k^{(34)}-\phi_k^{(34)})^2},\label{renormsite2energy}\nonumber
\eeqn
\begin{widetext}
\begin{eqnarray}
 \Delta_{k+1}^{(12)} & = & 
 \frac{\gamma_k^{(12)}\left(\gamma_k^{(34)}\Delta _k^{(13)}+2 \Delta _k^{(12)} \Delta _k^{(14)}\right)+2 \Delta _k^{(12)}\left(2\Delta
  _k^{(12)}\Delta _k^{(13)} +\gamma_k^{(34)}\Delta _k^{(23)} \right)}
  { \left({{(4{\Delta_k^{(12)}}^2+{\gamma_k^{(12)}}^2)(4{\Delta_k^{(12)}}^2+{\gamma_k^{(34)}}^2)}}\right)^{1/2}}
%  \Delta_{k+1}^{(12)} & = & 
%  \frac{\left( a b \Delta _k^{(13)}-2 a \Delta _k^{(23)} \Delta _k^{(12)}-2 b \Delta _k^{(14)} \Delta
%   _k^{(12)}+4 \Delta _k^{(13)} {\Delta _k^{(12)}}^2 \right)}
%   { \left({{(4{\Delta_k^{(12)}}^2+(\epsilon_k^{(12)}-\phi_k^{(12)})^2)(4{\Delta_k^{(12)}}^2+(\epsilon_k^{(34)}-\phi_k^{(34)})^2)}}\right)^{1/2}} 
,\label{renormcoupenergysimp}
  \nonumber\\
g_{k+1}^{(12)}(\textbf{q})&=&M(\textbf{q})\cos{\left(\frac{\textbf{d}_{k}\cdot\textbf{q}}{2}\right)}\left[\cos{\left(\frac{\textbf{d}_{k+1}\cdot\textbf{q}}{2}\right)}\left(\cos^2 {\theta_{k-}^{(12)}}-\cos^2{\theta_{k-}^{(34)}}\right)\right.%\nonumber\\
%& &~~~~~~~~~~~~~~~~~~~~~~~~~~~~~~~~~~~~~~~~~~~~~
\left.+\sin{\left(\frac{\textbf{d}_{k+1}\cdot\textbf{q}}{2}\right)}\left(\cos^2 {\theta_{k-}^{(12)}} +\cos^2 {\theta_{k-}^{(34)}}\right)\right],\label{grenormgendis}
\end{eqnarray}
\end{widetext}
where $\gamma_k^{(ij)}=\phi_k^{(ij)}-\epsilon_k^{(ij)}$.

\subsection{Parameterization and scale-based crossover}
Parameterizations of the transfer coupling $\Delta$ range from more exact methods \cite{frahmcke06,rozbicki08} to the simpler ideal dipole approximation (IDA) \cite{frahmcke06}. We implement the latter as $\Delta^{(ij)}=\kappa|\eta^{(i)}||\eta^{(j)}|/(4\pi\epsilon_T^{(ij)}{d^{(ij)}}^3)$, where $\eta^{(i),(j)}$ are the transition dipole moments of sites $i$ and $j$, and $\epsilon_T^{(ij)}$ the dielectric permittivity between them. $\kappa$ is a factor determined by the relative orientations of $\eta^{(i),(j)}$ and their connection vector. We set $\kappa=1$ for convenience, since our model accommodates arbitrary configurations. In real systems, $\eta^{(m)}$ varies if multiple chromophore types are present, and $\epsilon_T^{(ij)}$ may change with each coupling, depending on the multiscale geometry of the network and material composition of its surroundings (eg. protein embedded in lipid, suspended in aqueous fluid) \cite{dekker05, gilmore06}. We avoid an elaborate, arbitrary choice of composition and geometry, assuming a single chromophore type with transition dipole moment magnitude $\eta=4.5D$ \cite{damjanovic00}, and also uniform dielectric permittivity of $\epsilon_T=2.1\epsilon_0$, where $\epsilon_0$ is the vacuum permittivity, as for intracomplex couplings in LHCII \cite{linnanto06}. We take $\mu=1500kg/{m}^3$ for the mass density and $v=2500ms^{-1}$ for the speed of sound. $d_0$ is set at $10\AA$, based on the n-n spacings in LHCII. To our knowledge, the deformation potentials for chromophores in biological LHCs are not yet known. We estimate the deformation potential to be of order $D\approx0.1eV$ since this number gives physically plausible decoherence rates (Figure \ref{fig:deco}) and predicts transfer dynamics which fall narrowly inside the coherent regime at the 0th hierarchy level (Figure 3 in the main text), which is consistent with experimental findings for LHCs \cite{engel07, collini10, panit10, schlau09}.  

\begin{figure}
\includegraphics[angle=0,width=0.9\columnwidth]{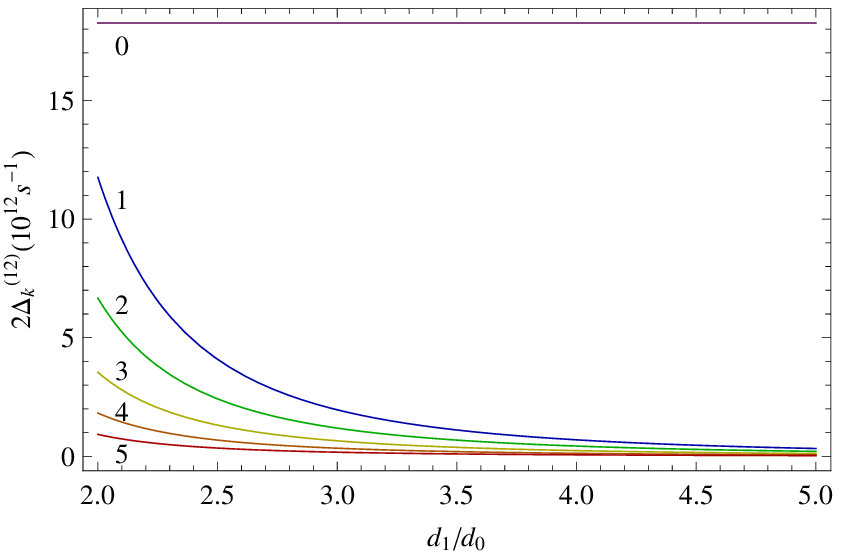}
\caption{\label{fig:transcoup} \textbf{Variations in the effective intradimer transfer couplings}, $2\Delta_k$, with degree of network clustering ($d_1/d_0$), for first six hierarchy levels $k$ (curves labelled).}
\end{figure}

An intuitive explanation for the scaling of the transfer dynamics is as follows. At the $k$th heirarchy level, the network geometry and surrounding dielectric set the transfer coupling rate $2\Delta_k$, which determines the value of the modulating function $W_k(2\Delta_k)$ by which the spectral density $J_k(2\Delta_k)$, and accordingly the decoherence rate $F_{k}(2\Delta_k)$, are modulated. The ratio of transfer rate to decoherence rate then determines whether the dynamics are coherent or incoherent. Unsurprisingly, $\Delta_k$ varies inversely with hierarchy level and degree of network clustering ($d_1/d_0$), though increased clustering curtails the relative reductions in $\Delta_k$ between levels (Figure \ref{fig:transcoup}). Counterintuitively however, $F_k(2\Delta_k)$ also varies inversely with level and clustering, with (for our chosen parameters) a large reduction between the $0$th and $1$st levels and smaller reductions thereafter (Figure \ref{fig:deco}). For large $k$, $2\Delta_k\ll v/d_k$ and $F_k(\omega)\approx Bk_BTd_k^2\omega^4/(6 v^2)$. Hence for our chosen geometry, $F_{k+1}(2\Delta_{k+1})\approx4(\Delta_{k+1}/\Delta_k)^4F_{k}(2\Delta_k)$. Therefore, for sufficiently large $k$, the decoherence rate vanishes.

\subsection{Acknowledgements}
We are grateful to the following people for discussions and comments: B. Hankamer, M. Landsberg, E. Knauth, I. Ross, M. Sarovar, S. Hoyer, B. Whaley, A. Ishizaki, T. Calhoun, G. Schlau-Cohen, N. Ginsberg, J. Dawlaty, G. Fleming, P. Rebentrost, L. Vogt, A. Perdomo, M. Mohseni, A. Aspuru-Guzik, A. Olaya-Castro, S. Jang, R. Pfeifer, P. Rohde, E. Cavalcanti, R. McKenzie. AKR thanks the Whaley and Fleming groups at UC Berkeley, and the Aspuru-Guzik group at Harvard, for hospitality. This work was supported by Australian Research Council grants CE110001013, FF0776191, DP1093287, DP0986352 and a Dan David Prize doctoral scholarship.

\end{document}